\documentclass[12pt]{article}

\usepackage{amsbsy}
\usepackage{amsmath}
\usepackage{amssymb}
\usepackage{amsthm}
\usepackage{authblk}
\usepackage{bbm}
\usepackage{color}
\usepackage{enumitem}
\usepackage{geometry}
\usepackage{graphicx}
\usepackage{natbib}
\usepackage{mathtools}
\usepackage{setspace}
\usepackage{enumerate}
\usepackage{natbib}

\usepackage{multirow}
\usepackage{booktabs}
\usepackage{times}
\usepackage{mathrsfs}
\usepackage{float}

\usepackage{booktabs}

\PassOptionsToPackage{hyphens}{url}
\usepackage[colorlinks = true,
            linkcolor = blue,
            urlcolor  = blue,
            citecolor = blue,
            anchorcolor = blue]{hyperref}
\usepackage{etoolbox}


\providecommand{\keywords}[1]
{
  \small	
  \textbf{\textit{Keywords:}} #1
}

\usepackage{bm}

\newtheorem{theorem}{Theorem}
\newtheorem{lemma}{Lemma}
\newtheorem{proposition}{Proposition}
\newtheorem{remark}{Remark}

\newtheorem{Condition}{Condition}
\newtheorem{Corollary}{Corollary}

\def\proof{{\it Proof.\ }}

\def\hat{\widehat}
\def\tilde{\widetilde}

\title{\LARGE\bf Maximum smoothed likelihood method for the combination of multiple diagnostic tests, with application to the ROC estimation}
\date{} 
 
\begin{document}
\def\spacingset#1{\renewcommand{\baselinestretch}%
{#1}\small\normalsize} 

\author{Fangyong Zheng$^1$,  Pengfei Li$^2$, and Tao Yu$^3$}

\date{}
\maketitle
\vspace{-0.36in}

 \begin{center}
 $^1$Fangyong Zheng is PhD Candidate,
  Department of Statistics and Data Science,
 National University of Singapore, Singapore, 117546\\
  (Email: \emph{e1117764@u.nus.edu})
  \\
  $^2$Pengfei Li is Professor,
  Department of Statistics and Actuarial Sciences,
  University of Waterloo,
  Waterloo, ON, Canada, N2L 3G1\\
  (Email: \emph{pengfei.li@uwaterloo.ca})\\
  $^3$Tao Yu is Associate Professor,
  Department of Statistics and Data Science,
 National University of Singapore, Singapore, 117546\\
  (Email: \emph{yu.tao@nus.edu.sg})
  \\
  \end{center}

\begin{abstract}
In medical diagnostics, leveraging multiple biomarkers can significantly improve classification accuracy compared to using a single biomarker. While existing methods based on exponential tilting or density ratio models have shown promise, their assumptions may be overly restrictive in practice. In this paper, we adopt a flexible semiparametric model that relates the density ratio of diseased to healthy subjects through an unknown monotone transformation of a linear combination of biomarkers. To enhance estimation efficiency, we propose a smoothed likelihood framework that exploits the smoothness in the underlying densities and transformation function. Building on the maximum smoothed likelihood methodology, we construct estimators for the model parameters and the associated probability density functions. We develop an effective computational algorithm for implementation, derive asymptotic properties of the proposed estimators, and establish procedures for estimating the receiver operating characteristic (ROC) curve and the area under the curve (AUC). Through simulation studies and a real-data application, we demonstrate that the proposed method yields more accurate and efficient estimates than existing approaches.
\end{abstract}


\spacingset{1.9}

\section{Introduction}

In medical diagnosis, multiple biomarkers are often collected for each subject. Each biomarker may carry partial and complementary information about the presence, severity, or progression of a disease. Relying on a single biomarker may lead to incomplete or inaccurate assessments, particularly when the disease is heterogeneous or affects multiple physiological pathways. Therefore, appropriately incorporating and combining multiple biomarkers has become increasingly important for improving diagnostic accuracy, enhancing risk stratification, and enabling more personalized medical decisions \citep{Kodosaki2024Combination, Kim2021ProteinPanel}. Effective combination strategies can exploit the joint distribution of biomarkers to construct powerful diagnostic tools, such as composite scores or multivariate ROC curves, that better distinguish between diseased and non-diseased individuals; see \citet{Ghosal2024_CutoffROC} and the references therein. 

Proposing the appropriate combination methods, establishing effective numerical algorithms, and exploring the underline theoretical properties have been popular research topics in the statistical community for decades. Assuming that the biomarkers follow a multivariate normal distribution, \cite{su1993linear} established the best linear combination approach. Based on the Neyman–Pearson lemma, \cite{mcintosh2002combining} showed that the density ratio between the diseased and healthy subjects maximizes the true positive rate at any fixed false positive rate, making it the gold standard for combining multiple biomarkers; see also \citet{eguchi2002class, copas2002overestimation}. Methods for combining biomarkers, established on the density ratio model, are available in the literature. \cite{qin2010best} considered an exponential tilting model, where they assumed that the log density ratio between the diseased and non-diseased populations is a combination of multiple diagnostic tests; specifically, the model is given by 
\begin{equation}\label{exp-tilting}
    \frac{f(x)}{g(x)}=\exp\left\{\alpha + \beta^\tau \xi(x) \right\},
\end{equation}
where $\alpha$ is a scalar parameter, $x\in \mathbb{R}^d$, $\beta$ is a $p$ dimensional parameter, $\xi(\cdot)$ is a known $p\times 1$ smooth vector function of $x$, $f(\cdot)$ and $g(\cdot)$ respectively denote the joint probability density functions (p.d.f.s) of the biomarker measurements from the diseased and healthy subjects. \cite{chen2016using} observed that the exponential function form assumed by \eqref{exp-tilting} could be restrictive and may not be well satisfied in practice; see Section \ref{Simulation Study} for examples. They proposed the semiparameteric model:
\begin{equation}\label{model_use-1}
    \frac{f(x)}{g(x)}=\phi(\beta^\tau \xi(x)),
\end{equation}
where $\phi(\cdot)$ is an unknown nondecreasing function, and an empirical likelihood based method for estimating the model parameters. More developments of methods for combining the biomarkers can be found in \cite{ma2007combining}, \citet{barreno2008optimal}, \citet{liu2013roc},  \citet{kim2013exploring}, \citet{HuangSanda2022_LinearBiomarker}, among others.  

Model~\eqref{model_use-1} is less restrictive than the exponential tilting model~\eqref{exp-tilting}, and can yield more accurate estimation of $\beta$ when the assumption underlying~\eqref{exp-tilting} is violated. Therefore, we adopt model~\eqref{model_use-1} in this paper. While the empirical likelihood method proposed in \cite{chen2016using} is compatible with this model, it does not exploit potential smoothness in the underlying densities $f(x)$ and $g(x)$, or in the monotonic transformation function $\phi(\cdot)$. In many practical applications, however, these functions are indeed smooth, and failing to leverage this property may lead to efficiency loss and suboptimal estimation. To address this issue, we propose a method under model~\eqref{model_use-1} that incorporates smoothness to enhance estimation accuracy of $f(\cdot)$, $g(\cdot)$, and $\beta$. Specifically, we adopt a smoothed likelihood approach, building on the maximum smoothed likelihood method introduced by \citet{eggermont1995maximum}. This technique parallels maximum likelihood estimation in parametric settings but is designed for nonparametric density estimation, preserving many desirable properties such as consistency and efficiency. The smoothed likelihood method has proven effective in a variety of challenging statistical problems, including inverse convolution~\cite{eggermont1995inverse}, smooth monotone and unimodal density estimation~\cite{eggermont2000maximum}, estimation of hazard rates and event-time distributions~\cite{groeneboom2010nonparametric}, density estimation in mixture models~\cite{levine2011, yu2019maximum}, interval-censored data~\cite{groeneboom2014maximum}, and two-sample problems with likelihood ratio ordering~\cite{yu2017density}. We build on these developments and incorporate the smoothed likelihood framework into our estimation procedure.

In this paper, assuming model \eqref{model_use-1}, we propose the smoothed likelihood and, based on it, define estimators \(\hat{\beta}\), \(\hat{g}(\cdot)\), and \(\hat{f}(\cdot)\) for the parameters \(\beta\), \(g(\cdot)\), and \(f(\cdot)\), respectively. We also establish and implement an efficient computational algorithm for estimating these parameters. Using these estimators, we further estimate the ROC curve and AUC. We rigorously derive the asymptotic properties of our parameter estimators, as well as those of the ROC and AUC estimators. Through extensive simulation studies, we demonstrate that our method achieves superior estimation accuracy compared to competing approaches. Finally, we illustrate the practical utility of our method by applying it to a real-data example.

The rest of the paper is organized as follows. Section~\ref{estimation} introduces the proposed maximum smoothed likelihood approach and defines the corresponding parameter estimators. Section~\ref{section-algorithm} presents the computational algorithm for parameter estimation. Section~\ref{section-Asymptotic Properties} establishes the asymptotic properties of the proposed estimators. Section~\ref{section-Estimating ROC and AUC} describes the estimation of the ROC curve and AUC, and investigates their asymptotic behavior. Section~\ref{Simulation Study} reports the results of simulation studies, and Section~\ref{data application} illustrates the proposed method using a pancreatic cancer dataset. Section~\ref{discussion} concludes the paper with a discussion. The technical conditions are given in the Appendix, while detailed technical derivations supporting the theoretical results in Sections~\ref{estimation}, ~\ref{section-Asymptotic Properties}, and~\ref{section-Estimating ROC and AUC} are included in the supplementary material.

\section{Maximum Smoothed Likelihood Estimation}\label{estimation}

Let $\{X_1, \ldots, X_n\}$ and $\{Y_1, \ldots, Y_m\}$ be the $d$-dimensional test results from diseased and healthy subjects, respectively. Assume that $X_1, \ldots, X_n$ are independent and identically distributed (i.i.d.) as a random vector $X$, and $Y_1, \ldots, Y_m$ are i.i.d. as $Y$. Denote by $F(x)$ and $f(x)$ the cumulative distribution function (c.d.f.) and p.d.f. of $X$, respectively; likewise, let $G(x)$ and $g(x)$ denote the c.d.f. and p.d.f. of $Y$. We assume the monotonic density ratio model as in equation~\eqref{model_use-1}. For presentational simplicity, and without loss of generality, we consider the following form:
\begin{equation}\label{model_use}
    \frac{f(x)}{g(x)} = \phi(\beta^\tau x),
\end{equation}
where $\phi(\cdot)$ is an unknown, nondecreasing function supported on $\mathbb{R}$, and thus $p = d$. To ensure the identifiability of the model, we assume that the first component of $\beta$ equals 1. The log-likelihood up to a constant coeffient is 
\begin{equation}\label{loglikelihood}
     \widetilde l_{n,m}(f,g)=\frac{1}{n+m}\left[\sum_{i=1}^{n} {\rm log} \left\{f(X_{i})\right\}+\sum_{j=1}^{m}{\rm log} \left\{g(Y_{j})\right\}\right],
\end{equation}
subject to \eqref{model_use}, and 
\begin{eqnarray}\label{eq-constraint-fg}
    f(x)\geq 0, \quad  g(x) \geq 0,  \quad \int_{\mathbb{R}^d}f(x) dx =\int_{\mathbb{R}^d}g(x) dx =1. 
\end{eqnarray}
This log-likelihood, however, is unbounded, as we can make $f(X_i)$ arbitrarily large for each $X_i$. The smoothed likelihood approach can be borrowed to tackle this unbounded problem; see \citet{yu2017density, yu2019maximum, levine2011}, and the references therein. Specifically, let \( N_Hf(x) \) be a nonlinear smoothing operator on a density \( f(\cdot) \), given by 
\[
N_Hf(x) = \exp\left\{\int_{\mathbb{R}^d} K_H(u - x) \, \log f(u) \, du\right\},
\]
where \( K_H(x) = |H|^{-1} K(H^{-1}x) \). Here, \( K(\cdot) \) is a symmetric multivariate p.d.f., serving as the kernel function; \( H \) is a $d\times d$ diagonal, positive definite bandwidth matrix. We replace $f$ and $g$ in the log-likelihood \eqref{loglikelihood} respectively with $N_{H_1}f$ and $N_{H_2}g$ to obtain the smoothed log-likelihood: 
\begin{equation}\label{smoothloglikelihood}
     l_{n,m}(f,g)=\frac{1}{n+m}\left[\sum_{i=1}^{n} {\rm log} \left\{ N_{H_1}f(X_{i}) \right\}+\sum_{j=1}^{m}{\rm log} \left\{N_{H_2}g(Y_{j})\right\}\right],
\end{equation}
where $H_1$ and $H_2$ are the bandwidth matrices, $H_1, H_2 \to 0$ as $n\to \infty$.

Let $\tilde{f}_n(x)=n^{-1}\sum_{i=1}^{n}K_{H_1}(x-X_i)$, $\tilde{g}_m(x)=m^{-1}\sum_{j=1}^{m}K_{H_2}(x-Y_j)$, and $\lambda = n/(n+m)$; for simplicity, we assume $\lambda \in(0,1)$ and is a constant. The smoothed log-likelihood $l_{n,m}(f,g)$ in \eqref{smoothloglikelihood} can be rewritten to be 
 \begin{equation}\label{rewrite smooth}
     l_{n,m}(f,g)=\lambda\int_{\mathbb{R}^d}\tilde{f}_{n}(x){\rm log}\left\{f(x)\right\}dx+(1-\lambda)\int_{\mathbb{R}^d}\tilde{g}_{m}(x){\rm log}\left\{g(x)\right\}dx.
 \end{equation}
 With the reparameterization:
 \begin{equation*}
    \theta(\beta^\tau x)=\frac{\lambda f(x)}{\lambda f(x)+(1-\lambda)g(x)}=\frac{\lambda \phi(\beta^\tau x)}{\lambda \phi(\beta^\tau x)+(1-\lambda)}, \quad \psi(x)=\lambda f(x)+(1-\lambda)g(x),
\end{equation*}
giving 
\begin{equation}\label{Giving}
    f(x)=\frac{\theta(\beta^\tau x)\psi(x)}{\lambda}\,,\qquad g(x)=\frac{\left\{1-\theta(\beta^\tau  x)\right\} \psi(x)}{1-\lambda},
\end{equation}
we can write the smoothed log-likelihood function in \eqref{rewrite smooth} to be
$$l_{n,m}(f,g)=l_{n,m,1}(\theta,\beta)+l_{n,m,2}(\psi)+\text{constant},$$where
\begin{eqnarray}
    l_{n,m,1}(\theta,\beta)&=&\int_{\mathbb{R}^d}\lambda \widetilde{f}_{n}(x){\rm log}\, \theta(\beta^{\tau}x)dx+\int_{\mathbb{R}^d}(1-\lambda)\widetilde{g}_{m}(x){\rm log}\,\{1-\theta(\beta^{\tau}x)\}dx \label{eq-def-l-n-m-1}\\
    l_{n,m,2}(\psi)&=&\int_{\mathbb{R}^d}\left\{\lambda \widetilde{f}_{n}(x)+(1-\lambda)\widetilde{g}_{m}(x)\right\}{\rm log}\,\psi(x)dx. \nonumber 
\end{eqnarray}
Hereafter, we shall take $\theta(\cdot), \beta, \psi(\cdot)$ as the parameters of interest; based on \eqref{model_use} and \eqref{eq-constraint-fg}, our proposed estimators for $\theta(\cdot), \beta, \psi(\cdot)$ are the maximizers of $l_{n,m,1}(\theta,\beta)$ and $l_{n,m,2}(\psi)$, subject to 
\begin{equation} \label{eq-constraints}
\begin{aligned}
&\psi(\cdot) \geq 0,\quad \int_{\mathbb{R}^d}\psi(x) dx=1; \quad \int_{\mathbb{R}^d} \psi(x)\theta(\beta^\tau x)dx=1\,;\\
    &\,\theta(\cdot)\in [0,1] \text{ and is non-decreasing}. \\
\end{aligned}
\end{equation}

The estimation of $\psi(\cdot)$ and $(\theta(\cdot), \beta)$ can be carried out based on $l_{n,m,2}(\psi)$ and $l_{n,m,1}(\theta, \beta)$ separately. Referring to \cite{eggermont2001maximum}, page 122, the maximizer of $l_{n,m,2}(\psi)$ is given by 
\begin{eqnarray} \label{def-psi-hat}
\hat{\psi}_{n,m}(x)=\lambda \tilde{f}_n(x)+(1-\lambda)\tilde{g}_m(x),
\end{eqnarray}
which serves as the estimator for $\psi(\cdot)$. The estimators for $\theta(\cdot), \beta$ are defined to be
\begin{eqnarray} \label{def-theta-beta-hat}
    (\hat \theta(\cdot), \hat \beta) = \arg\max_{\theta \in \Theta, \beta \in \textbf{B}}l_{n,m,1}(\theta,\beta),
\end{eqnarray}
where $\Theta = \left\{\theta(\cdot): \theta(\cdot)\in [0,1] \text{ and is non-decreasing} \right\}$ and $\textbf{B} = \{1\}\times \textbf{B}_{-1}$ are the parameter spaces for $\theta(\cdot)$ and $\beta$. The optimization problem \eqref{def-theta-beta-hat} can be achieved by a profiling procedure as follows. For each $\beta \in \textbf{B}$, let 
\begin{eqnarray} 
    \theta_\beta (\cdot) = \arg\max_{\theta \in \Theta} l_{n,m,1}(\theta,\beta). \label{profile-beta} 
\end{eqnarray}
Theorem \ref{Guarantee fhat pdf} below ensures that the optimization problem \eqref{profile-beta} is equivalent to a continuous weighted isotonic regression. 

\begin{theorem}\label{Guarantee fhat pdf}
Assume Condition \ref{Condition-1} in the Appendix. For each $\beta \in \textbf{B}$, consider $\theta_\beta(\cdot)$ defined in \eqref{profile-beta}.
\begin{itemize}
    \item[(a)] Equivalently, $\theta_\beta(\cdot)$ is also the solution of a continuous weighted isotonic regression: 
\begin{eqnarray}
\theta_\beta(\cdot) = \arg\min_{\theta \in \Theta}\int_{\mathbb{R}^d}\left\{\frac{\lambda \tilde{f}_{n}(x)}{\hat{\psi}_{n,m}(x)}-\theta(\beta^\tau  x)\right\}^2 \hat{\psi}_{n,m}(x)dx. \label{eq-weighted-isotonic-reg}
\end{eqnarray}
\item[(b)] Consider $\hat \psi_{n,m}(\cdot)$ given by \eqref{def-psi-hat}. We have $$\int_{\mathbb{R}^d} \hat{\psi}_{n,m}(x)\theta_{\beta}(\beta^\tau x)dx=\lambda.$$
\end{itemize}
\end{theorem}
Part (a) of Theorem \ref{Guarantee fhat pdf} illustrates that $\theta_\beta(\cdot)$ is the solution of a continuous weighted isotonic regression \citep{groeneboom2010generalized}, and can be readily solved numerically; the details are given in Section \ref{section-algorithm}. Subsequently, the profile likelihood is given by 
\begin{eqnarray}
l^{*}(\beta)=\int_{\mathbb{R}^d}\lambda \widetilde{f}_{n}(x){\rm log}\, \theta_{\beta}(\beta^{\tau}x)dx+\int_{\mathbb{R}^d}(1-\lambda)\widetilde{g}_{m}(x){\rm log}\,\{1-\theta_{\beta}(\beta^{\tau}x)\}dx, \label{eq-weighted-isotonic-reg-profile}
\end{eqnarray}
and thus
\begin{eqnarray*}
    \hat \beta = \arg\max_{\beta \in \textbf{B}} l^{*}(\beta) \quad \text{and} \quad \hat \theta(\cdot) = \theta_{\hat\beta}(\cdot). 
\end{eqnarray*}
Based on Part(b) of Theorem \ref{Guarantee fhat pdf}, we observe that $\hat \psi(\cdot), \hat \theta(\cdot), \hat \beta$ are the maximizers of $l_{m,n,1}(\theta, \beta)$, $l_{n,m,2}(\psi)$ subject to \eqref{eq-constraints}. Consequently, the estimators for $f(\cdot)$ and $g(\cdot)$ can be obtained by 
\begin{eqnarray*}
    \hat{f}(x)=\hat{\theta}(\hat{\beta}^{\tau}x)\hat{\psi}(x)/\lambda,\qquad \hat{g}(x)=\left\{1-\hat{\theta}(\hat{\beta}^{\tau}x)\right\} \hat{\psi}(x)/(1-\lambda). 
\end{eqnarray*}

\section{Computational Algorithm} \label{section-algorithm}

We develop a numerical algorithm for computing the proposed estimators in Section~\ref{estimation}. The challenges lie in the fact that the objective functions in \eqref{eq-weighted-isotonic-reg} and \eqref{eq-weighted-isotonic-reg-profile} are formed of $d$-dimensional integrals, which can be computationally intensive to evaluate directly. To avoid this multiple dimensional integration, we propose a Monte Carlo estimation procedure.  

First of all, with data $X_1, \ldots, X_n$, $Y_1, \ldots, Y_m$, we need the bandwidth $H_1$ and $H_2$ to compute $\tilde f_n(x)$ and $\tilde g_m(x)$ through 
\begin{eqnarray*}
\tilde{f}_n(x)=\frac{1}{n}\sum _{i=1}^{n}K_{H_1}(x-X_i)\,,\quad\tilde{g}_m(x)=\frac{1}{m}\sum_{j=1}^{m}K_{H_2}(x-Y_j). 
\end{eqnarray*}
We adopt the plug-in method of \cite{wand1994multivariate} to obtain $H_1$ and $H_2$ based on $X_1, \ldots, X_n$ and $Y_1, \ldots, Y_m$ respectively. 

We then generate $N$ samples, $Z_1, \ldots, Z_N$ from $\tilde f_n(\cdot)$, and $N$ samples, $Z_{N+1}, \ldots, Z_{2N}$ from $\tilde g_m(\cdot)$. Specifically, to  generate a random variable $Z$ from $\tilde f_n(\cdot)$, we can use the following steps: 
\begin{itemize}
    \item Step 1: Draw an $X$ from $X_1, \ldots, X_n$ with equal probability.
    \item Step 2: Draw a $U$ from $K(\cdot)$. 
    \item Step 3: Let $Z = X + H_1 U$. 
\end{itemize}
We can check that given $X_1, \ldots, X_n$, $Z \sim \tilde f_n(x)$. In particular, the conditional c.d.f. of $Z$ is given by 
\begin{eqnarray*}
&&P(Z \leq x|X_1, \ldots, X_n)=P(X+H_1\cdot U \leq x|X_1, \ldots, X_n)\\
&=&\frac{1}{n}\sum_{i=1}^{n}P\left( U \leq H_1^{-1}(x-X_i)|X_i\right)=\frac{1}{n}\sum_{i=1}^{n}\mathcal{K}(H_1^{-1}(x-X_i)),
\end{eqnarray*}
which is the corresponding conditional c.d.f. for $\tilde f_n(x)$, where $\mathcal{K}(\cdot)$ denotes the corresponding c.d.f. of $K(\cdot)$. Similarly, conditional on $Y_1, \ldots, Y_m$, we can generate $Z\sim \tilde g_m(x)$. Subsequently, the objective function in \eqref{eq-weighted-isotonic-reg} can be estimated by
\begin{eqnarray}
&&\int_{\mathbb{R}^d}\left\{\frac{\lambda \tilde{f}_{n}(x)}{\hat{\psi}_{n,m}(x)}-\theta(\beta^\tau  x)\right\}^2 \hat{\psi}_{n,m}(x)dx \label{Monte Calo Approx}  \\
&\approx& \frac{1}{N}\lambda\sum_{i=1}^{N}\left\{\frac{\lambda \tilde{f}_{n}(Z_i)}{\hat{\psi}_{n,m}(Z_i)}-\theta(\beta^\tau  Z_{i})\right\}^2+\frac{1}{N}(1-\lambda)\sum_{i=N+1}^{2N}\left\{\frac{\lambda \tilde{f}_{n}(Z_i)}{\hat{\psi}_{n,m}(Z_i)}-\theta(\beta^\tau  Z_{i}) \right\}^2. \nonumber 
\end{eqnarray}
To ensure the approximation accuracy, we can set $N$ to be a large value; in particular, in our numerical studies, we have used $N = 10,000$. 
For each $\beta \in \text{B}$, finding $\theta\in \Theta$ that maximizes \eqref{Monte Calo Approx} can be achieved by applying the standard pool adjacent violators (PAVA) algorithm \citep{Ayer1955}; the details are given in the following steps. 
\begin{itemize}
    \item Step 1: sort $\left\{\beta^\tau  Z_i\right\}_{i=1}^{2N}$ from smallest to largest, and denote 
    \begin{eqnarray*}
        \beta^\tau Z_{(1)} \leq \ldots \leq  \beta^\tau Z_{(2N)}. 
    \end{eqnarray*}
    \item Step 2: accordingly, the objective function \eqref{Monte Calo Approx} can be written to be 
    \begin{eqnarray}
        \frac{1}{N}\sum_{i=1}^{2N}\left\{\frac{\lambda \tilde{f}_{n}(Z_{(i)})}{\hat{\psi}_{n,m}(Z_{(i)})}-\theta(\beta^\tau  Z_{(i)})\right\}^2 w_{(i)}, \label{Monte Calo Approx-1}
    \end{eqnarray}
    where $w_{(i)} = \lambda$ if $Z_{(i)}$ is from $\tilde f_n(\cdot)$, and $w_{(i)} = 1-\lambda$, otherwise. 

    \item Step 3: the available PAVA algorithms, e.g., \texttt{pava()} function in R, can be applied to solve \eqref{Monte Calo Approx-1} numerically. 
\end{itemize}

Finally, the profile likelihood in \eqref{eq-weighted-isotonic-reg-profile} can be approximated by 
\begin{equation}\label{beta fun}
    \frac{1}{N}\lambda\sum_{i=1}^{N}\,{\rm log}\,\left\{\theta_{\beta}(\beta^\tau  Z_i)\right\}+\frac{1}{N}(1-\lambda)\sum_{i=N+1}^{2N}{\rm log}\, \left\{1-\theta_{\beta}(\beta^\tau  Z_i)\right\},
\end{equation}
and can be readily solved by the \texttt{optim()} function in R. 

\section{Asymptotic Properties} \label{section-Asymptotic Properties}

In this section, we explore the asymptotic properties of our estimators. Let $F_0(\cdot)$, $ f_0(\cdot)$, $G_0(\cdot)$, $g_0(\cdot)$, $\theta_0(\cdot)$, and  $\beta_0$ respectively denote their corresponding counter parts, and let 
\begin{eqnarray*}
    d(\theta_1, \beta_1; \theta_2, \beta_2)=\left[\int \left\{\theta_1(\beta_1^\tau   x)-\theta_2(\beta_2^\tau x)\right\}^{2}\,d\Psi_0(x)\right]^{1/2}, 
\end{eqnarray*}
where $\Psi_0(x)=\lambda F_0(x)+(1-\lambda)G_0(x)$. Furthermore, we denote 
\begin{eqnarray*}
    \alpha_n = \max_{H\in \{H_1, H_2\}; f \in \{f_0, g_0\}}\int_{\mathbb{R}^d} |K_{H}*f(x)-f(x)| dx, 
\end{eqnarray*}
where $K*f(x)$ denotes the convolution of functions $K$ and $f$: $K*f(x) = \int_{\mathbb{R}^d} K(x-y)f(y) dy$.
Based on the standard results in kernel smoothing, we have $\alpha_n \to 0$ as $n\to \infty$. In particular, we have the following lemma.

\begin{lemma} \label{lemma-K-f}
Let $H = \text{diag}(h_1, \ldots, h_d)$, such that for each $i=1,\ldots, d$, $h_i\to 0$, and $K(\cdot)$ satisfy Condition \ref{Condition-1} in the Appendix. Moreover, let $f(x)$ be a p.d.f. defined on $x \in \mathbb{R}^d$, satisfying
\begin{eqnarray*}
    \int |D^\alpha f(x)|dx<\infty,
\end{eqnarray*}
where $\alpha = (\alpha_1, \ldots, \alpha_d)^\tau$ such that $\alpha_1 + \ldots + \alpha_d  = 2$, and $D^\alpha f=\frac{\partial^{\alpha_1+\cdots+\alpha_d}f}{\partial x_1^{\alpha_1}\cdots \partial x_d^{\alpha_d}}$. Then,
\begin{eqnarray*}
    \int \left|K_H*f(x)-f(x)\right|dx = O\left\{\left(\max\limits_{1\leq i\leq d} h_i\right)^2\right\}. 
\end{eqnarray*}
\end{lemma}

We have the following theorem. 

\begin{theorem}\label{thetabeta}
    Assume Conditions \ref{Condition-1}--\ref{Condition-5} in the Appendix. We have  
\begin{itemize}
    \item[(a)] $d(\hat \theta, \hat \beta; \theta_0, \beta_0)=O_p(n^{-1/3})\vee O_p(\sqrt{\alpha_n}), $
    \item[(b)] $\hat{\beta}-\beta_0=O_p(n^{-1/3})\vee O_p(\sqrt{\alpha_n})$. 
\end{itemize}
\end{theorem}
Based on Theorem \ref{thetabeta}, we are able to establish the convergence of $\hat f(\cdot)$ and $\hat g(\cdot)$. 
\begin{Corollary} \label{corollary-1}
Assume Conditions \ref{Condition-1}--\ref{Condition-6} in the Appendix. Let 
\begin{eqnarray*}
    \eta_n = \max\left\{|H_1|, \ |H_2|\right\},
\end{eqnarray*}
where $|H| = h_1 \cdots h_d$\ for\ $H = \text{diag}(h_1,\ldots, h_d)$. Assuming that $n \eta_n \to \infty$, we have
\begin{eqnarray*}
\int\left\{\hat{f}(x)-f_0(x)\right\}^2dx&=&O_p\left(n^{-2/3}\right)\vee O_p\left(\alpha_n^2\right)\vee O_p\left(1/(n\eta_n)\right) \\ \int\left\{\hat{g}(x)-g_0(x)\right\}^2dx&=&O_p\left(n^{-2/3}\right)\vee O_p\left(\alpha_n^2\right)\vee O_p\left(1/(n\eta_n)\right).
\end{eqnarray*}
\end{Corollary}

\begin{remark} \label{remark-opt-rate}
    Consider the case where $H_1 = H_2 = \text{diag}(h, \ldots, h)$, and assume that the conditions of Lemma~\ref{lemma-K-f} are satisfied. Under this setup, the convergence rate in Theorem~\ref{thetabeta} becomes
\[
O_p\left(n^{-1/3}\right) \vee O_p(\sqrt{\alpha_n}) = O_p\left(n^{-1/3}\right) \vee O_p(h),
\]
which simplifies to $O_p\left(n^{-1/3}\right)$ if we choose $h = O_p\left(n^{-1/3}\right)$. This represents the best convergence rate achievable by our estimators based on Theorem~\ref{thetabeta}.

Similarly, for $d \geq 2$, the $l_2$ convergence rate of $\hat{f}(\cdot)$ and $\hat{g}(\cdot)$, based on Corollary~\ref{corollary-1}, is given by
\[
\sqrt{O_p\left(n^{-2/3}\right) \vee O_p\left(\alpha_n^2\right) \vee O_p\left(1/(n \eta_n)\right)} = O_p\left(n^{-1/3}\right) \vee O_p(h^2) \vee O_p\left(1/\sqrt{n h^d}\right),
\]
with the optimal rate being $O_p\left\{n^{-2/(4+d)}\right\}$, attained when $h = O_p\left\{n^{-1/(4+d)}\right\}$.
\end{remark}

\section{Estimation for ROC and AUC} \label{section-Estimating ROC and AUC}

The ROC curve and AUC have been popularly used statistical tools and measurements in many scientific areas, especially in the medical research. In the framework of Section \ref{estimation}, the combined biomarkers are respectively $U=\beta^\tau X, \tilde U = \beta^\tau Y$. Denote $F_C(u)=P\left(U \leq u\right)$ and $G_C(u)=P\left(\tilde U \leq u\right)$. The ROC curve and AUC are given by \citet{pepe2003statistical}:
\begin{eqnarray*}
    {\rm ROC}_C(s)&=&1-F_C\left(G_C^{-1}(1-s)\right) \\
    {\rm AUC}_C &=& \int_{0}^{1}{\rm ROC}_C(s)ds,
\end{eqnarray*}
for $s\in [0,1]$. Furthermore, when $\beta = \beta_0$, we denote $F_{C, 0}(u)$, $G_{C,0}(u)$, ${\rm ROC}_{C,0}(s)$, and ${\rm AUC}_{C,0}$ accordingly; in this case, ${\rm ROC}_{C,0}(s)$ is the optimal ROC curve \citep{eguchi2002class, mcintosh2002combining}. With our estimates of $\beta$, $f(\cdot)$, and $g(\cdot)$ in Section \ref{estimation}, we can estimate $F_C(\cdot)$ and $G_C(\cdot)$ by
\begin{eqnarray*}
    \hat{F}_C(u)&=&\int I\left(\hat{\beta}^\tau x \leq u\right)\hat{f}(x)\,dx, \\ \hat{G}_C(u)&=&\int I\left(\hat{\beta}^\tau x \leq u\right)\hat{g}(x)\,dx. 
\end{eqnarray*}
As a consequence, the corresponding ROC curve estimate and the AUC estimate are given by 
\begin{eqnarray}
\widehat{{\rm ROC}}_C(s)&=&1-\hat{F}_C\left(\hat{G}_C^{-1}(1-s)\right)\,\,\text{for each $s \in (0,1)$}, \label{ROC-estimate}\\
\widehat{{\rm AUC}}_C&=&\int_{0}^{1}\widehat{{\rm ROC}}_C(s)ds. \nonumber 
\end{eqnarray}
We observe that $\widehat{{\rm ROC}}_C(\cdot)$ satisfies the following proposition. 
\begin{proposition} \label{proposition-ROC}
Assume Condition \ref{Condition-1} in the Appendix. The ROC estimate given by \eqref{ROC-estimate} satisfies $\widehat{{\rm ROC}}_C(0) = 0, \widehat{{\rm ROC}}_C(1) = 1$, and $\widehat{{\rm ROC}}_C(s)$ is a non-decreasing and concave function for $s \in [0,1]$. 
\end{proposition}
The following theorem establishes the asymptotic properties of these estimates.
\begin{theorem}\label{converge of CDF}
    Assume Conditions \ref{Condition-1}--\ref{Condition-5} in the Appendix.  We have 
\begin{itemize}
    \item[(a)]  $\sup \limits_{u\in \mathbb{R}}\left|\hat{F}_C(u)-F_{C,0}(u)\right|=O_p(n^{-1/3})\vee O_p(\sqrt{\alpha_n})\,,$
    
    $\sup\limits_{u\in \mathbb{R}}\left|\hat{G}_C(u)-G_{C,0}(u)\right|=O_p(n^{-1/3})\vee O_p(\sqrt{\alpha_n});$
    \item[(b)] $\sup\limits_{s \in (0,1)}\left|\widehat{\rm{ROC}}_C(s)-\rm{ROC}_{C,0}(s)\right|=O_p(n^{-1/3})\vee O_p(\sqrt{\alpha_n});$
    \item[(c)] $\widehat{\rm{AUC}}_C={\rm{AUC}}_{C,0}+O_p\left(n^{-1/3}\right)\vee O_p(\sqrt{\alpha_n}).$
\end{itemize}
\end{theorem}

Based on Theorem \ref{converge of CDF} and the discussion in Remark \ref{remark-opt-rate}, the best convergence rate for all these estimators is $O_p\left(n^{-1/3}\right)$, attained when $h = O_p\left(n^{-1/3}\right)$.  

\section{Simulation Study} \label{Simulation Study}

\subsection{Settings}

In our numerical study, we compare our method (named ``Our") with three competitive methods: (1) the empirical likelihood based method by \cite{chen2016using} (named ``Empirical"); (2) the exponential tilting method by \cite{qin2010best} (named ``Exp Tilting"); (3) the method established on maximizing AUC by \cite{ma2007combining} (named ``MH"). 

We consider two simulation examples: Examples 1 and 2. For Example 1, we set $d = 2$. Let $f_1(\cdot), f_2(\cdot)$, and $f_3(\cdot)$ be the p.d.f.s of lognormal(0,1), lognormal(1,1), and lognormal(4,1), respectively. We simulate $X_1, \ldots, X_n$ to be i.i.d., with  
\begin{eqnarray*}
    X_i \sim f(x_1, x_2) = \rho f_1(x_1)f_1(x_2) + (1-\rho) f_2(x_1)f_3(x_2); \label{eq-simu-1}
\end{eqnarray*}
and simulate $Y_1, \ldots, Y_m$ to be i.i.d. with 
\begin{eqnarray*}
    Y_i \sim g(x_1, x_2) = (1-\rho) f_1(x_1)f_1(x_2) + \rho f_2(x_1)f_3(x_2), \label{eq-simu-2}
\end{eqnarray*}
where different values of $\rho$ are considered in our numerical studies. Note that the p.d.f. of the $\text{lognormal}(\mu, \sigma)$ distribution is given by 
\begin{eqnarray*}
\frac{1}{x \sigma \sqrt{2\pi}}\exp\left\{-\frac{(\log x -\mu)^2}{2\sigma^2}\right\}. \label{eq-simu-3}
\end{eqnarray*} 
Thus, $U(x_1, x_2) \equiv \log \left\{\frac{f_1(x_1)f_1(x_2)}{f_2(x_1) f_3(x_2)}\right\}$ is of the structure: 
\begin{eqnarray}
    U(x_1, x_2) = \log\left\{\frac{f_1(x_1)f_1(x_2)}{f_2(x_1) f_3(x_2)}\right\} = \alpha+\beta_1 \log x_1+\beta_2 \log x_2. \label{eq-simu-4}
\end{eqnarray}
Furthermore, 
\begin{eqnarray}
    \frac{f(x_1, x_2)}{g(x_1, x_2)}=\frac{\rho\exp(U)+(1-\rho)}{(1-\rho)\exp(U)+\rho}. \label{eq-simu-5}
\end{eqnarray}

In practice, $f_1(x_1) f_1(x_2)$ and $f_2(x_1) f_3(x_2)$ may represent the joint p.d.f.s of the biomarkers for diseased and healthy subjects based on the ground truth, respectively. The quantity $1 - \rho$ represents the proportion of individuals in the identified diseased group who are actually healthy, and likewise, the proportion in the identified healthy group who are actually diseased. Thus, $\rho = 1$ corresponds to the ideal scenario in which all diseased and healthy subjects are correctly identified. In this case, $\log \left(f(x_1, x_2)/g(x_1, x_2)\right) = U(x_1, x_2)$, so the model assumptions for our method and all competing methods are fully satisfied. Similarly, $\rho \in (0.5, 1)$ corresponds to the scenario where a proportion $1 - \rho$ of subjects in both the diseased and healthy groups are misclassified. Based on \eqref{eq-simu-4} and \eqref{eq-simu-5}, the model assumptions for the ``Our'', ``Empirical'', and ``MH'' methods are satisfied, whereas those for the ``Exp Tilting'' method are violated.

For Example 2, we consider $d = 3$. Let $g_1(\cdot) \sim N(0,1)$, $g_2(\cdot) \sim \text{Gamma}(2.5, 4)$, $g_3(\cdot) \sim N(1,1)$, $g_4(\cdot) \sim N(4.5,1)$, $g_5(\cdot) \sim \text{Gamma}(2,4)$. We simulate $X_1, \ldots, X_n$, and $Y_1,\ldots, Y_m$ by
\begin{eqnarray*}
    X_i &\sim& f(x_1,x_2,x_3) = \rho g_1(x_1)g_1(x_2)g_2(x_3) + (1-\rho) g_3(x_1) g_4(x_2)g_5(x_3) \\
    Y_i &\sim& g(x_1,x_2,x_3) = (1-\rho) g_1(x_1)g_1(x_2)g_2(x_3) + \rho g_3(x_1) g_4(x_2)g_5(x_3). 
\end{eqnarray*}
Similarly to Example 1, $g_1(x_1)g_1(x_2)g_2(x_3)$ and $g_3(x_1) g_4(x_2)g_5(x_3)$ may represent the joint p.d.f.s of the biomarkers for diseased and healthy subjects, respectively, and satisfy 
\begin{eqnarray*}
    U(x_1, x_2, x_3) = \log \left\{\frac{g_1(x_1)g_1(x_2)g_2(x_3)}{g_3(x_1) g_4(x_2)g_5(x_3)} \right\} = \alpha+\beta_1x_1+\beta_2x_2+\beta_3\log x_3. 
\end{eqnarray*}
The quantity $1 - \rho$ represents the proportion of individuals in the identified diseased group who are actually healthy, and likewise, the proportion in the identified healthy group who are actually diseased.

For both examples, we consider the combinations of setting:  $\rho = 1, 0.9, 0.8$; $n = m = 300$ and $n=m = 600$. 

\subsection{Results of $\beta$ estimates}

In this section, we compare the $\beta$ estimates from different methods. For comparison purpose, we regularize $\|\beta\|_2 =1$, where $\|\cdot\|_2$ denotes the $l_2$ norm in the Euclidean space. Tables \ref{Table for lognormal case compare} and \ref{Table for Joint distribution of gamma and normal case compare} summarize the results 
 of Examples 1 and 2 from different methods over 1000 replicates; all reported values are multiplied by 1000 to provide a clearer view. From both tables, we observe that for both examples, our method consistently yields the lowest or near-lowest mean square errors (MSEs) and standard deviations (SDs) across different sample sizes and levels of model misspecification. Its performance remains stable even when $\rho$ is small, indicating strong robustness to model misspecification. The Empirical and Exp tilting methods show moderate accuracy, but their performance deteriorates under stronger misspecification. The MH method generally performs the worst, with high variability and large MSE under strong misspecification and smaller samples. This suggests limited robustness and efficiency in such scenarios.

\begin{table}[H]
\centering
\caption{Comparison of $\beta$ estimates for Example 1. All reported values have been multiplied by 1000.}
\label{Table for lognormal case compare}
\resizebox{1\textwidth}{!}{
\begin{tabular}{rrr|rrr|rrr|rrr|rrr}
\toprule
\multicolumn{3}{c|}{} & \multicolumn{3}{c|}{Our} & \multicolumn{3}{c|}{Empirical} & \multicolumn{3}{c|}{Exp Tilting} & \multicolumn{3}{c}{MH} \\
$n=m$ & $\rho$ & $\beta$ & Bias & SD & MSE & Bias & SD & MSE & Bias & SD & MSE & Bias & SD & MSE \\
\midrule

 & 1.0 & $\beta_1$ & -2.87 & 60.97 & 3.73 & 6.70 & 99.24 & 9.90 & 7.81 & 82.64 & 6.89 & 6.40 & 93.49 & 8.78 \\
 & 1.0 & $\beta_2$ & 2.77 & 15.67 & 0.25 & 3.75 & 25.11 & 0.64 & 1.82 & 20.84 & 0.44 & 3.23 & 24.11 & 0.60 \\
\addlinespace
300 & 0.9 & $\beta_1$ & -13.57 & 85.40 & 7.48 & 2.39 & 151.81 & 23.05 & 0.48 & 116.93 & 13.67 & 16.45 & 302.67 & 91.88 \\
 & 0.9 & $\beta_2$ & 7.56 & 23.63 & 0.62 & 12.24 & 41.57 & 1.88 & 7.43 & 30.45 & 0.98 & 48.24 & 86.44 & 9.80 \\
 \addlinespace
 & 0.8 & $\beta_1$ & -20.65 & 120.48 & 14.94 & 4.43 & 221.76 & 49.20 & 3.54 & 162.80 & 26.52 & 8.62 & 371.52 & 138.10 \\
 & 0.8 & $\beta_2$ & 13.58 & 34.81 & 1.40 & 26.93 & 67.40 & 5.27 & 13.85 & 43.68 & 2.10 & 78.52 & 111.42 & 18.58 \\
\midrule

 & 1.0 & $\beta_1$ & -0.05 & 47.03 & 2.21 & 5.85 & 73.38 & 5.42 & 1.14 & 55.87 & 3.12 & -0.66 & 63.10 & 3.98 \\
 & 1.0 & $\beta_2$ & 1.22 & 11.74 & 0.14 & 1.50 & 18.08 & 0.33 & 1.43 & 14.13 & 0.20 & 2.36 & 16.39 & 0.27 \\
\addlinespace
600 & 0.9 & $\beta_1$ & -7.77 & 64.70 & 4.25 & -0.81 & 112.50 & 12.66 & -1.21 & 80.78 & 6.53 & 9.77 & 249.32 & 62.25 \\
 & 0.9 & $\beta_2$ & 4.29 & 17.30 & 0.32 & 7.22 & 30.24 & 0.97 & 3.90 & 21.22 & 0.47 & 32.59 & 68.70 & 5.78 \\ \addlinespace
 & 0.8 & $\beta_1$ & -15.07 & 90.58 & 8.43 & 4.37 & 156.60 & 24.54 & 1.71 & 112.18 & 12.59 & 14.58 & 317.64 & 101.11 \\
 & 0.8 & $\beta_2$ & 8.48 & 25.43 & 0.72 & 12.55 & 42.40 & 2.00 & 6.55 & 30.43 & 0.97 & 54.23 & 91.37 & 11.29 \\
\bottomrule
\end{tabular}
}
\end{table}

\begin{table}[H]
\centering
\caption{Comparison of $\beta$ estimates for Example 2. All reported values have been multiplied by 1000.}
\label{Table for Joint distribution of gamma and normal case compare}
\resizebox{0.95\textwidth}{!}{
\begin{tabular}{rrr|rrr|rrr|rrr|rrr}
\toprule
\multicolumn{3}{c|}{} & \multicolumn{3}{c|}{Our} & \multicolumn{3}{c|}{Empirical} & \multicolumn{3}{c|}{Exp Tilting} & \multicolumn{3}{c}{MH} \\
$n=m$ & $\rho$ & $\beta$ & Bias & SD & MSE & Bias & SD & MSE & Bias & SD & MSE & Bias & SD & MSE \\
\midrule

 & 1.0 & $\beta_1$ & 3.56 & 78.60 & 6.19 & 8.71 & 119.61 & 14.38 & 0.81 & 102.21 & 10.45 & 10.72 & 112.00 & 12.66 \\
 & 1.0 & $\beta_2$ & 7.84 & 22.33 & 0.56 & 20.58 & 37.46 & 1.83 & 15.24 & 31.56 & 1.23 & 17.95 & 34.75 & 1.53 \\
 & 1.0 & $\beta_3$ & -8.74 & 108.79 & 11.91 & 6.77 & 161.35 & 26.08 & -6.66 & 140.30 & 19.73 & 4.88 & 155.68 & 24.26 \\
\addlinespace
 & 0.9 & $\beta_1$ & -22.28 & 107.80 & 12.12 & 5.86 & 181.67 & 33.04 & 2.74 & 124.59 & 15.53 & 45.81 & 324.61 & 107.47 \\
300 & 0.9 & $\beta_2$ & 24.89 & 37.56 & 2.03 & 48.89 & 63.44 & 6.42 & 22.13 & 38.03 & 1.94 & 145.94 & 109.66 & 33.32 \\
 & 0.9 & $\beta_3$ & 12.49 & 147.44 & 21.89 & 7.36 & 237.52 & 56.47 & -0.84 & 163.97 & 26.89 & -22.34 & 408.47 & 167.35 \\
\addlinespace
 & 0.8 & $\beta_1$ & -32.01 & 143.08 & 21.50 & 21.06 & 242.85 & 59.42 & 6.79 & 165.61 & 27.47 & 36.85 & 371.64 & 139.47 \\
 & 0.8 & $\beta_2$ & 44.25 & 56.28 & 5.13 & 83.17 & 88.32 & 14.72 & 39.50 & 56.59 & 4.76 & 190.87 & 125.83 & 52.26 \\
 & 0.8 & $\beta_3$ & 15.86 & 204.59 & 42.11 & 34.12 & 296.67 & 89.17 & 0.46 & 217.60 & 47.35 & -22.83 & 446.53 & 199.91 \\
\midrule

 & 1.0 & $\beta_1$ & 3.51 & 60.96 & 3.73 & 5.71 & 88.66 & 7.89 & 3.25 & 70.81 & 5.03 & 4.16 & 80.65 & 6.52 \\
 & 1.0 & $\beta_2$ & 4.13 & 16.12 & 0.28 & 10.11 & 25.40 & 0.75 & 6.11 & 19.33 & 0.41 & 8.78 & 23.52 & 0.63 \\
 & 1.0 & $\beta_3$ & -5.42 & 81.67 & 6.70 & 2.99 & 113.18 & 12.82 & -6.49 & 95.87 & 9.23 & -3.52 & 111.59 & 12.46 \\
\addlinespace
 & 0.9 & $\beta_1$ & -19.27 & 79.12 & 6.63 & 13.88 & 134.12 & 18.18 & 6.90 & 90.76 & 8.28 & 29.96 & 297.77 & 89.56 \\
600 & 0.9 & $\beta_2$ & 16.49 & 25.40 & 0.92 & 25.83 & 43.21 & 2.53 & 10.67 & 25.37 & 0.76 & 119.55 & 97.42 & 23.78 \\
 & 0.9 & $\beta_3$ & 11.01 & 116.88 & 13.78 & 15.26 & 178.70 & 32.17 & -3.03 & 123.76 & 15.33 & -18.26 & 368.05 & 135.80 \\
\addlinespace
 & 0.8 & $\beta_1$ & -32.89 & 108.13 & 12.77 & 32.40 & 184.40 & 35.05 & 10.30 & 122.86 & 15.20 & 37.55 & 339.43 & 116.62 \\
 & 0.8 & $\beta_2$ & 29.58 & 39.22 & 2.41 & 44.62 & 63.19 & 5.98 & 19.38 & 35.22 & 1.62 & 161.16 & 110.67 & 38.22 \\
 & 0.8 & $\beta_3$ & 15.34 & 156.65 & 24.77 & 8.57 & 240.24 & 57.79 & -8.25 & 164.28 & 27.06 & -16.49 & 421.61 & 178.03 \\
\bottomrule
\end{tabular}
}
\end{table}

\subsection{Results of ROC and AUC estimates}

In this section, we compare the ROC and AUC estimates from our method with those from other methods. For the ROC estimates, we use the averaged $L_2$ distance between $\widehat{\rm{ROC}}_C(\cdot)$ and  $\rm{ROC}_{C,0}(\cdot)$ over 1000 replications as the comparison criteria, in particular, this $L_2$ distance is defined to be:
\begin{eqnarray*}
    L_2\left(\widehat{\rm{ROC}}_C, \rm{ROC}_{C,0}\right)=\left[\int_{0}^{1}\left\{\widehat{\rm{ROC}}_C(s)-\rm{ROC}_{C,0}(s)\right\}^2ds\right]^{1/2}.
\end{eqnarray*}
For the AUC estimates, we use the relative bias (RB) and MSE over 1000 replications as the comparison criteria. The RB in percentage is defined to be: 
\begin{eqnarray*}
RB(\%)=\frac{\widehat{\text{AUC}}_C- \text{AUC}_{C,0}}{\text{AUC}_{C,0}}\times 100. 
\end{eqnarray*}

Tables~\ref{ROC-AUC-table-example-1} presents the $L_2$ distances for ROC estimation, and the RB (\%) and MSE for AUC estimation, across different sample sizes and misclassification rates in Example 1. Similarly, Tables~\ref{ROC-AUC-table-example-2} presents the corresponding results for Example 2. Based on these tables, we have the following observations.

Our method consistently achieves the lowest or among the lowest $L_2$ distances across all scenarios, indicating more accurate ROC curve estimation. The Empirical method generally performs second best but produces larger $L_2$ errors than our method when $\rho$ decreases. The Exp Tilting method tends to have higher $L_2$ distances, especially at lower $\rho$ values, which is expected since lower $\rho$ indicates greater model misspecification for this method. The MH method has comparable performance to the Empirical method. 

For AUC estimation, our method achieves the smallest MSE and absolute RB in most cases, indicating the least bias and highest precision among all methods. The MH method also yields small biases, comparable to our method, but tends to produce slightly larger MSEs, suggesting less precise AUC estimates. In contrast, the Empirical and Exp Tilting methods exhibit relatively larger RB, and the MSE values often exceed or match those of the MH method.

In summary, our method outperforms the other approaches in terms of both ROC curve and AUC estimation accuracy, maintaining low bias and error even under moderate misclassification and smaller sample sizes. The Empirical and MH methods provide reasonable performance but are less robust to misclassification errors. The Exp Tilting method consistently exhibits the largest errors and bias, indicating it may be less suitable in settings with misclassified groups.

\begin{table}[H]
\centering
\caption{Example 1: $L_2$ distances ($\times 1000$) for ROC estimation and RB (\%) and MSE ($\times 1000$) for AUC estimation.}
\label{combined-table}
\resizebox{1\textwidth}{!}{
\begin{tabular}{ll|rrr|rrr|rrr|rrr}
\toprule
\multicolumn{2}{c|}{} 
& \multicolumn{3}{c|}{Our} 
& \multicolumn{3}{c|}{Empirical} 
& \multicolumn{3}{c|}{Exp Tilting} 
& \multicolumn{3}{c}{MH} \\
\cmidrule(lr){3-5} \cmidrule(lr){6-8} \cmidrule(lr){9-11} \cmidrule(lr){12-14}
$n=m$ & $\rho$ 
& $L_2$ & RB & MSE 
& $L_2$ & RB & MSE 
& $L_2$ & RB & MSE 
& $L_2$ & RB & MSE \\
\midrule
    & 1.0  & 6.97 & -0.17 & $<0.01$ & 4.86 & 0.03 & $<0.01$ & 3.32 & -0.02 & $<0.01$ & 4.91 & 0.00 & $<0.01$ \\
300 & 0.9  & 27.53 & -0.09 & 0.17 & 33.50 & 1.20 & 0.27 & 49.73 & 0.90 & 0.24 & 32.59 & 0.11 & 0.19 \\
    & 0.8  & 26.77 & 0.22 & 0.31 & 34.09 & 2.21 & 0.61 & 44.32 & 1.14 & 0.42 & 32.33 & 0.41 & 0.35 \\
\hline
    & 1.0  & 5.68 & -0.14 & $<0.01$ & 3.65 & 0.01 & $<0.01$ & 2.31 & -0.02 & $<0.01$ & 3.75 & -0.01 & $<0.01$ \\
600 & 0.9  & 20.24 & -0.16 & 0.09 & 23.62 & 0.73 & 0.13 & 47.02 & 0.85 & 0.15 & 23.31 & -0.09 & 0.10 \\
    & 0.8  & 19.50 & 0.23 & 0.16 & 24.21 & 1.56 & 0.31 & 41.00 & 1.19 & 0.26 & 23.08 & 0.27 & 0.18 \\
\hline
\end{tabular} \label{ROC-AUC-table-example-1}
}
\end{table}

\begin{table}[H]
\centering
\caption{Example 2: $L_2$ distances ($\times 1000$) for ROC estimation, and RB (\%) and MSE ($\times 1000$) for AUC estimation.}
\label{combined-3D}
\resizebox{1\textwidth}{!}{
\begin{tabular}{ll|rrr|rrr|rrr|rrr}
\toprule
\multicolumn{2}{c|}{} 
& \multicolumn{3}{c|}{Our} 
& \multicolumn{3}{c|}{Empirical} 
& \multicolumn{3}{c|}{Exp Tilting} 
& \multicolumn{3}{c}{MH} \\
\cmidrule(lr){3-5} \cmidrule(lr){6-8} \cmidrule(lr){9-11} \cmidrule(lr){12-14}
$n=m$ & $\rho$ 
& $L_2$ & RB & MSE 
& $L_2$ & RB & MSE 
& $L_2$ & RB & MSE 
& $L_2$ & RB & MSE \\
\midrule
    & 1.0 & 2.52 & -0.06 & $<0.01$ & 2.44 & -0.02 & $<0.01$ & 1.87 & -0.03 & $<0.01$ & 2.37 & -0.02 & $<0.01$ \\
300 & 0.9 & 28.54 & 0.47 & 0.20 & 35.73 & 1.53 & 0.35 & 48.13 & 1.42 & 0.34 & 33.82 & 0.52 & 0.22 \\
    & 0.8 & 28.49 & 0.80 & 0.39 & 36.79 & 2.57 & 0.75 & 43.40 & 1.69 & 0.55 & 33.79 & 0.89 & 0.42 \\
\hline
    & 1.0 & 2.13 & -0.06 & $<0.01$ & 1.92 & -0.02 & $<0.01$ & 1.31 & -0.04 & $<0.01$ & 1.84 & -0.03 & $<0.01$ \\
600 & 0.9 & 20.94 & 0.26 & 0.10 & 24.57 & 0.96 & 0.16 & 44.99 & 1.32 & 0.23 & 24.12 & 0.19 & 0.10 \\
    & 0.8 & 20.03 & 0.48 & 0.18 & 25.03 & 1.62 & 0.32 & 38.94 & 1.49 & 0.31 & 23.60 & 0.39 & 0.18 \\
\hline
\end{tabular} \label{ROC-AUC-table-example-2}
}
\end{table}

\section{Application}\label{data application}

Pancreatic cancer is among the most lethal forms of cancer, with a five-year survival rate of less than $10\%$ once diagnosed. However, early detection significantly improves the chances of survival. Unfortunately, pancreatic cancer often remains asymptomatic until it has reached an advanced stage and metastasized. Therefore, a reliable diagnostic test capable of identifying pancreatic cancer at an early stage could be of great clinical value.

In \cite{debernardi2020combination}, the authors collected a series of urinary biomarkers from three groups of patients: healthy controls; patients with non-cancerous pancreatic conditions such as chronic pancreatitis; and patients with pancreatic ductal adenocarcinoma. The sample sizes for these groups were 183, 143, and 83, respectively. The urinary proteins measured included LYVE1, REG1B, and TFF1. We proceed to combine the test results of LYVE1, REG1B, and TFF1. To this end, we adopt the following semiparametric monotonic density ratio model:
\[
\frac{f(x)}{g(x)} = \phi\left(\beta_1 (100 \times \text{LYVE1}) + \beta_2 (\text{REG1B}) + \beta_3 (\text{TFF1})\right),
\]
where multiplying LYVE1 by 100 serves to bring its scale in line with the other biomarkers.

We use the data collected in the same institution-Barts Pancreas Tissue Bank, London, UK (BRTB). We view the  patients with non-cancerous pancreatic conditions as healthy people, thus the non-diseased group sample containing the Healthy control group plus the non-cancerous pancreatic group, i.e $n=83$, $m=326$. We report the coefficient estimates and their corresponding standard errors (SEs) based on 300 bootstrap estimates; we have regularized $\left\|\beta\right\|_2=1$ for comparison purposes. The results are summarized in Table \ref{Scenarios I}.

We observe that the estimates from the
four methods are slightly different; the difference between our
method and the ``Empirical" method is smallest. Our method leads
to smaller SEs than other methods. In contrast to the simulation example, we are unable to obtain the biases and MSEs of
the $\beta$ estimates in this real-data example. However, based on our observations in the simulation studies, we conjecture that the data of patients with non-cancerous pancreatic condition cause the misspecification of exponential tilting model. 

\begin{table}[H]
\centering
\caption{Estimates of the coefficients and their standard errors SE($\times 10$) }
\label{Scenarios I}
\begin{tabular}{@{}ccccccccc@{}}
\toprule
& \multicolumn{2}{c}{Our} & \multicolumn{2}{c}{Empirical} & \multicolumn{2}{c}{Exp Tilting} & \multicolumn{2}{c}{MH} \\
\cmidrule(lr){2-3} \cmidrule(lr){4-5} \cmidrule(lr){6-7} \cmidrule(lr){8-9}
Parameter & Estimate & SE & Estimate & SE & Estimate & SE & Estimate & SE \\
\midrule
$\beta_1$ & 0.90 & 1.01 & 0.90 & 1.30 & 0.74 & 1.50 & 0.84 & 2.04 \\
$\beta_2$ & 0.43 & 1.26 & 0.44 & 1.82 & 0.67 & 1.60 & 0.54 & 2.57 \\
$\beta_3$ & 0.02 & 0.52 & 0.02 & 0.54 & 0.02 & 0.63 & 0.05 & 0.73 \\
\bottomrule
\end{tabular}
\end{table}

\begin{figure}
    \centering
    \includegraphics[width=1\linewidth]{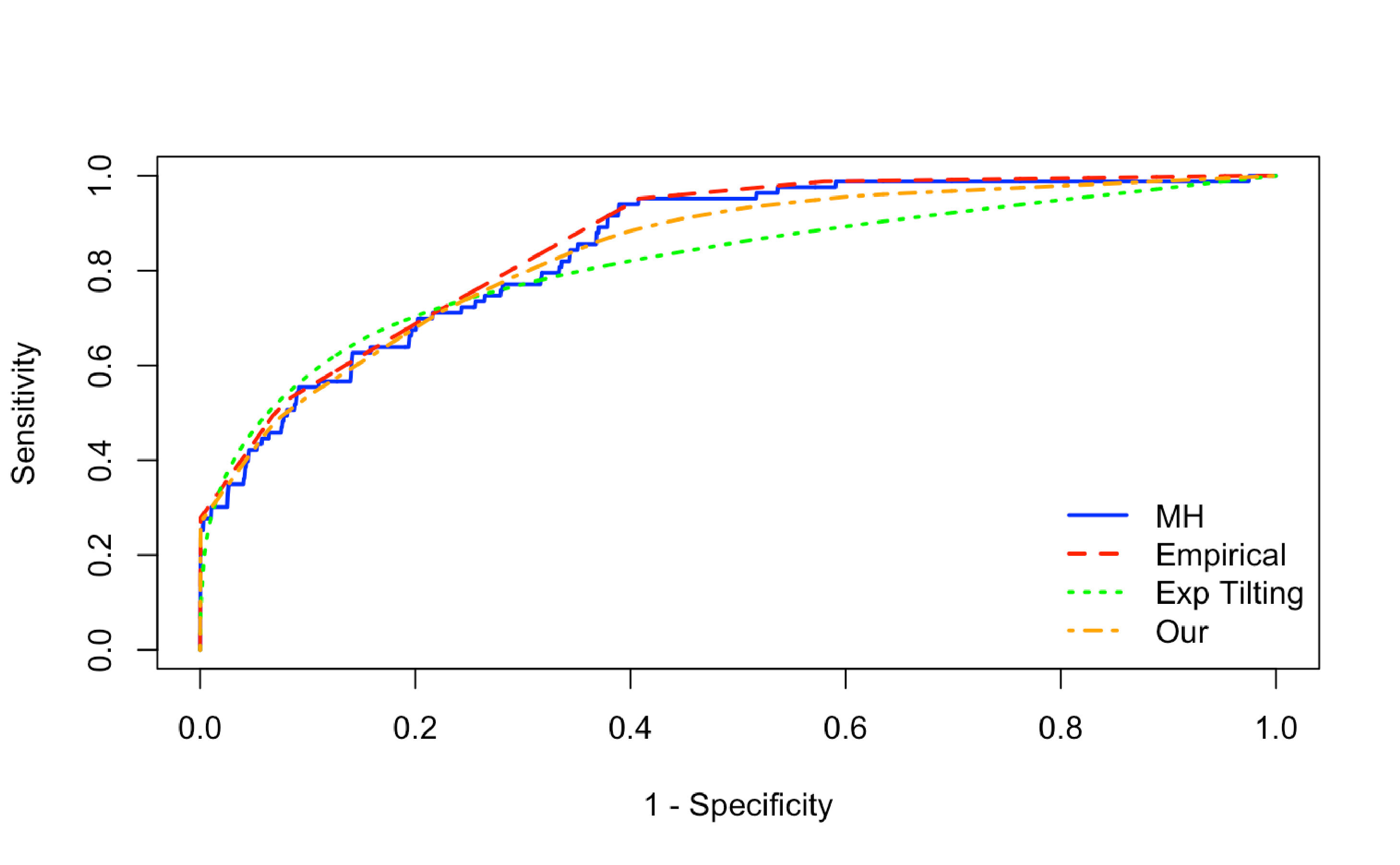}
    \caption{ROC estimates}
    \label{fig:roc of real data}
\end{figure}

Based on the $\beta$ estimates, we construct the ROC curves using the four methods, as shown in Figure~\ref{fig:roc of real data}, and summarize the corresponding AUC estimates in Table~\ref{Table for AUC of real data application}. The Our, Empirical, and Exp Tilting methods yield similar ROC curves and AUC values, indicating general agreement. The MH method shows slightly more deviation in both the ROC curve and AUC estimate. Unlike in the simulation study, however, we cannot assess the accuracy of these estimates against the true values due to the lack of ground truth in the real data.

\begin{table}[H]
\centering
\caption{Estimate and SE ($\times 1000$) of four methods for estimating the AUC.}
\label{Table for AUC of real data application}
\resizebox{0.8\textwidth}{!}{
\begin{tabular}{cc|cc|cc|cc}
\toprule
\multicolumn{2}{c|}{Our} & \multicolumn{2}{c|}{Empirical} & \multicolumn{2}{c|}{Exp Tilting} & \multicolumn{2}{c}{MH} \\
\cmidrule(lr){1-2} \cmidrule(lr){3-4} \cmidrule(lr){5-6} \cmidrule(lr){7-8}
Estimate & SE & Estimate & SE & Estimate & SE & Estimate & SE \\
\midrule
0.84 & 20.30 & 0.86 & 18.07 & 0.81 & 23.53 & 0.85 & 21.37 \\
\bottomrule
\end{tabular}
}
\end{table}

\section{Discussion} \label{discussion}
This paper introduces a maximum smoothed likelihood approach for combining multiple diagnostic tests under the semiparametric density ratio model defined in \eqref{model_use-1}. The proposed method accommodates the flexibility and smoothness of nonparametric components, leading to improved accuracy in parameter estimation and enhanced robustness against model misspecification.

We develop an efficient algorithm, establish the asymptotic properties of the parameter estimators, and derive estimators for the ROC curve and AUC along with their theoretical guarantees. Through numerical studies, we demonstrate that the proposed method consistently yields more accurate and stable estimates for both the combination coefficients and ROC/AUC measures, particularly in settings with model misspecification.

Despite its strengths, the performance of the proposed method could benefit from improved bandwidth selection for kernel smoothing. Additionally, it may face computational challenges when the number of biomarkers $d$ is large, and the current convergence rates may not be optimal. Future research could focus on automating bandwidth selection, enhancing scalability to high-dimensional settings, and establishing stronger theoretical guarantees.

\section*{Appendix: Regularity Conditions} 

We impose the following regularity conditions to establish our asymptotic results. They are not necessarily the weakest possible.

\begin{Condition} \label{Condition-1}
    $K(\cdot)$ is a symmetric p.d.f. with compact support in $\mathbb{R}^d$, and $\sup_{t\in \mathbb{R}^d} K(t) \lesssim 1$, where ``$\lesssim$" means ``smaller than, up to a universal constant". 
\end{Condition}

\begin{Condition} \label{Condition-3}
 $\beta \in \{1\} \times \text{\textbf{B}}_{-1}$ and $\text{\textbf{B}}_{-1}$ is a compact subspace of $R^{d-1}$
\end{Condition}

\begin{Condition} \label{Condition-2}
$H_1 = \text{diag}(h_{1,1}, \ldots, h_{1,d})$, $H_2 = \text{diag}(h_{2,1}, \ldots, h_{2,d})$. They satisfy: 
\begin{itemize}
    \item for each $s = 1,2; t = 1,\ldots, d$, $h_{s,t} \to 0$ as $n\to \infty$. 
    \item for any $s_1, s_2 \in \{1,2\}, t_1, t_2 \in \{1,\ldots, d\}$, 
    $$0 < \liminf_{n\to \infty } h_{s_1, t_1}/h_{s_2, t_2} \leq \limsup_{n\to \infty} h_{s_1, t_1}/h_{s_2, t_2} < C$$
    for $C$ being a universal constant.  
\end{itemize}
\end{Condition}

\begin{Condition} \label{Condition-4}
    For $\beta_1, \beta_2 \in \textbf{B}$ and $u\in \mathbb{R}^d$, there exists a universal constant $C$, not relying on $u$, such that $$\int \left|I\left(\beta_1^\tau x \leq u\right) -I\left(\beta_2^\tau x\leq u\right) \right|\,d P(x) \leq C \left\|\beta_1-\beta_2 \right\|_1,$$
    for $P = F_0, G_0$, and $\mathcal{K}$. 
\end{Condition}

\begin{Condition} \label{Condition-added-1}
There exists an $\eta_0 >0$, such that 
\begin{eqnarray*}
    \inf_{\beta: \|\beta-\beta_0\|_2 \leq \eta_0} \lambda_2 \left[ \text{var} \left\{ E\left( X|\beta^\tau X \right) - X \right\} \right] >0,
\end{eqnarray*}
where $\lambda_2(A)$ denotes the second minimum eigenvalue of matrix $A$.  
\end{Condition}

\begin{Condition}\label{Condition-5}
$\theta_0(\cdot)$ is second-order continuously differentiable with bounded second-order derivations, and there exist universal constants $0<c_1<c_2<1$, such that for any $x\in \mathbb{R}^d$, $c_1<\theta_0(\beta^\tau_0 x)<c_2$. Furthermore, for any $\beta \in \textbf{B}$ and $\theta(\cdot)$ being a monotonically increasing function such that $\theta(\beta^\tau x) = \theta_0(\beta_0^\tau x)$ for all $x$, then $\beta = \beta_0$ and $\theta(\cdot) = \theta_0(\cdot)$. 
\end{Condition}

\begin{Condition} \label{Condition-6}
Both $f_0(x)$ and $g_0(x)$ are bounded, and have bounded second order derivatives for $x \in \mathbb{R}^d$. 

\end{Condition}

\begin{remark} \label{remark-condition-4}
    With Condition \ref{Condition-4}, we have 
    \begin{eqnarray*}
        \int \left|I\left(\tilde \beta_1^\tau z \leq u\right) -I\left(\tilde \beta_2^\tau z\leq u\right) \right|\,d P(z) \leq  C\left\|\tilde \beta_1-\tilde \beta_2 \right\|_1, 
    \end{eqnarray*}
    where $\tilde \beta_1 = (\beta_1^T, \beta_{1,1}^T)^T$, $\tilde \beta_2 = (\beta_2^T, \beta_{2,1}^T)^T$, with $\beta_1, \beta_{1,1}, \beta_2, \beta_{2,1} \in \textbf{B}$, $z = (x^T,t^T)^T \in {\mathbb{R}^{2d}}$, and $P(z) = F_0(x)\mathcal{K}(t)$ or $P(z) = G_0(x)\mathcal{K}(t)$. 
\end{remark}
\proof This remark follows by:
\begin{eqnarray*}
&&\int \left|I\left(\tilde \beta_1^\tau z \leq u\right) -I\left(\tilde \beta_2^\tau z\leq u\right) \right|\,d P(z)\\
&=& \int\int \left|I\left(\beta_1^\tau x \leq u - \beta_{1,1}^\tau t\right) -I\left(\beta_2^\tau x\leq u - \beta_{2,1}^\tau t\right) \right|\,d F_0(x) d\mathcal{K}(t) \\
&\leq & \int \int \left|I\left(\beta_1^\tau x \leq u - \beta_{1,1}^\tau t\right) -I\left(\beta_2^\tau x\leq u - \beta_{1,1}^\tau t\right) \right|\,d F_0(x) d \mathcal{K}(t) \\
&& + \int \int \left|I\left(\beta_2^\tau x \leq u - \beta_{1,1}^\tau t\right) -I\left(\beta_2^\tau x\leq u - \beta_{2,1}^\tau t\right) \right|\,d F_0(x) d\mathcal{K}(t) \\
&\leq& C\|\beta_1-\beta_2\|_1 + \int \int \left|I\left(\beta_{1,1}^\tau t \leq u - \beta_2^\tau x\right) -I\left(\beta_{2,1}^\tau t\leq u - \beta_2^\tau x\right) \right|\, d\mathcal{K}(t)d F_0(x) \\
&\leq & C\|\beta_1-\beta_2\|_1 + C\|\beta_{1,1}-\beta_{2,1}\|_1 = C\|\tilde \beta_1 - \tilde \beta_2\|_1. 
\end{eqnarray*}

\begin{center}
    \large SUPPLEMENTARY MATERIAL
\end{center}
The supplementary material contains the technical details for the theoretical results in Sections \ref{estimation}, \ref{section-Asymptotic Properties}, and \ref{section-Estimating ROC and AUC}. 

\begin{center}
    \large DATA AVAILABILITY
\end{center}
The real pancreatic cancer data analysed in  Section \ref{data application} are downloadable from the following URL:
\begin{center}
{https://www.kaggle.com/datasets/johnjdavisiv/urinary-biomarkers-for-pancreatic-cancer}
\end{center}

\bibliographystyle{asa}
\bibliography{ref.bib}

@article{HuangSanda2022_LinearBiomarker,
  author       = {Huang, Yijian and Sanda, Martin G.},
  title        = {Linear biomarker combination for constrained classification},
  journal      = {The Annals of Statistics},
  volume       = {50},
  number       = {5},
  pages        = {2793--2815},
  year         = {2022},
  month        = oct,
  doi          = {10.1214/22-AOS2210},
  url          = {https://projecteuclid.org/euclid.aos/1666151654},
}

@article{Ghosal2024_CutoffROC,
  author       = {Ghosal, Soutik},
  title        = {Impact of Methodological Assumptions and Covariates on the Cutoff Estimation in {ROC} Analysis},
  journal      = {Biometrical Journal},
  volume       = {67},
  number       = {3},
  pages        = {e70053},
  year         = {2025},
  month        = {Apr},
  doi          = {10.1002/bimj.70053},
  pmid         = {40289475},
  pmc          = {PMC12035111},
  url          = {https://www.ncbi.nlm.nih.gov/pmc/articles/PMC12035111/},
}

@article{Kim2021ProteinPanel,
  author       = {
    Kim, Yoseop and 
    Yeo, Injoon and 
    Huh, Iksoo and 
    Kim, Jaenyeon and 
    Han, Dohyun and 
    Jang, Jin‑Young and 
    Kim, Youngsoo
  },
  title        = {Development and Multiple Validation of the Protein Multi‑marker Panel for Diagnosis of Pancreatic Cancer},
  journal      = {Clinical Cancer Research},
  year         = {2021},
  volume       = {27},
  number       = {8},
  pages        = {2236--2245},
  month        = apr,
  doi          = {10.1158/1078-0432.CCR-20-3929},
  url          = {https://doi.org/10.1158/1078-0432.CCR-20-3929},
  note         = {All seven authors included; first three (Kim, Y.; Yeo, I.; Huh, I.) contributed equally},
}

@article{Kodosaki2024Combination,
  author       = {Kodosaki, Eleftheria and Watkins, W John and Loveless, Sam and Kreft, Karim L and Richards, Aidan and Anderson, Valerie and Hurler, Lisa and Robertson, Neil P and Zelek, Wioleta M and Tallantyre, Emma C and others},
  title        = {Combination protein biomarkers predict multiple sclerosis diagnosis and outcomes},
  journal      = {Journal of Neuroinflammation},
  year         = {2024},
  volume       = {21},
  number       = {1},
  pages        = {52},
  doi          = {10.1186/s12974-024-03036-4},
  url          = {https://doi.org/10.1186/s12974-024-03036-4},
}

@article{groeneboom2014maximum,
  author    = {Groeneboom, Piet},
  title     = {Maximum Smoothed Likelihood Estimators for the Interval Censoring Model},
  journal   = {Annals of Statistics},
  year      = {2014},
  volume    = {42},
  number    = {5},
  pages     = {2092--2137},
  doi       = {10.1214/14-AOS1232},
  mrnumber  = {3262478}
}

@article{groeneboom2010nonparametric,
  author    = {Groeneboom, Piet and Jongbloed, Geurt and Witte, Bianca I.},
  title     = {Maximum Smoothed Likelihood Estimation and Smoothed Maximum Likelihood Estimation in the Current Status Model},
  journal   = {Annals of Statistics},
  year      = {2010},
  volume    = {38},
  number    = {1},
  pages     = {352--387},
  doi       = {10.1214/09-AOS716},
  mrnumber  = {2589324}
}

@article{eggermont2000maximum,
  author    = {Eggermont, Paul P. B. and LaRiccia, Vincent N.},
  title     = {Maximum Likelihood Estimation of Smooth Monotone and Unimodal Densities},
  journal   = {Annals of Statistics},
  year      = {2000},
  volume    = {28},
  number    = {4},
  pages     = {922--947},
  doi       = {10.1214/aos/1015952000},
  mrnumber  = {1792794}
}

@article{eggermont1995inverse,
  author    = {Eggermont, Paul P. B. and LaRiccia, Vincent N.},
  title     = {Maximum Smoothed Likelihood Density Estimation for Inverse Problems},
  journal   = {Annals of Statistics},
  year      = {1995},
  volume    = {23},
  number    = {1},
  pages     = {199--220},
  doi       = {10.1214/aos/1176324460},
  mrnumber  = {1331664}
}

@article{kim2013exploring,
  author       = {Kim, Yong-Sik and Jang, Min-Kyu and Park, Chan-Young and Song, Hyun-Jung and Kim, Jin-Dong},
  title        = {Exploring Multiple Biomarker Combination by Logistic Regression for Early Screening of Ovarian Cancer},
  journal      = {International Journal of Bio-Science and Bio-Technology},
  year         = {2013},
  volume       = {5},
  number       = {1},
  pages        = {67--76}
}

@inproceedings{barreno2008optimal,
  author       = {Barreno, Marco and Cardenas, Alvaro A. and Tygar, J. D.},
  title        = {Optimal ROC Curve for a Combination of Classifiers},
  booktitle    = {Advances in Neural Information Processing Systems (NIPS)},
  year         = {2008},
  volume       = {20},
  pages        = {57--64}
}

@article{liu2013roc,
  author       = {Liu, Danping and Zhou, Xiao-Hua},
  title        = {ROC Analysis in Biomarker Combination With Covariate Adjustment},
  journal      = {Academic Radiology},
  year         = {2013},
  volume       = {20},
  number       = {7},
  pages        = {874--882},
  doi          = {10.1016/j.acra.2013.02.003}
}

@article{Ayer1955,
	author = {Ayer, M and Brunk, H. D and Ewing, G. M and Reid, W. T and Silverman, E},
	journal = {The Annals of Mathematical Statistics},
	pages = {641-647},
	title = {An empirical distribution function for sampling with incomplete information},
	volume = {26},
	year = {1955}}

@article{levine2011,
	author = {M. Levine and D. R. Hunter and D. Chauveau},
	journal = {Biometrika},
	pages = {403-416},
	title = {Maximum smoothed likelihood for multivariate mixtures},
	volume = {98},
	year = {2011}}

@article{su1993linear,
  title={Linear combinations of multiple diagnostic markers},
  author={Su, JQ and Liu, JS},
  journal={Journal of the American Statistical Association},
  volume={88},
  pages={1350--1355},
  year={1993}
}

@article{eguchi2002class,
  title={A class of logistic-type discriminant functions},
  author={Eguchi, S and Copas, J},
  journal={Biometrika},
  volume={89},
  pages={1--22},
  year={2002}
}

@article{copas2002overestimation,
  title={Overestimation of the receiver operating characteristic curve for logistic regression},
  author={Copas, J and Corbett, P},
  journal={Biometrika},
  volume={89},
  pages={315--331},
  year={2002}
}

@article{mcintosh2002combining,
  title={Combining several screening tests: optimality of the risk score},
  author={McIntosh, MW and Pepe, MS},
  journal={Biometrics},
  volume={58},
  pages={657--664},
  year={2002}
}

@article{qin2010best,
  author = {Qin, J. and Zhang, B.},
  title = {Best Combination of Multiple Diagnostic Tests for Screening Purposes},
  journal = {Statistics in Medicine},
  volume = {29},
  pages = {2905--2919},
  year = {2010}
}

@article{chen2016using,
  author = {Chen, B. and Li, P. and Qin, J. and others},
  title = {Using a Monotonic Density Ratio Model to Find the Asymptotically Optimal Combination of Multiple Diagnostic Tests},
  journal = {Journal of the American Statistical Association},
  volume = {111},
  pages = {861--874},
  year = {2016}
}

@article{eggermont1995maximum,
  author = {Eggermont, P. P. B. and LaRiccia, V. N.},
  title = {Maximum Smoothed Likelihood Density Estimation},
  journal = {Nonparametric Statistics},
  volume = {4},
  pages = {211--222},
  year = {1995}
}

@article{yu2017density,
  author = {Yu, T. and Li, P. and Qin, J.},
  title = {Density Estimation in the Two-Sample Problem with Likelihood Ratio Ordering},
  journal = {Biometrika},
  volume = {104},
  pages = {141--152},
  year = {2017}
}

@article{groeneboom2010generalized,
  author = {Groeneboom, P. and Jongbloed, G.},
  title = {Generalized Continuous Isotonic Regression},
  journal = {Statistics \& Probability Letters},
  volume = {80},
  pages = {248--253},
  year = {2010}
}

@article{debernardi2020combination,
  title={A combination of urinary biomarker panel and PancRISK score for earlier detection of pancreatic cancer: A case--control study},
  author={Debernardi, Silvana and O’Brien, Harrison and Algahmdi, Asma S and Malats, Nuria and Stewart, Grant D and Plje{\v{s}}a-Ercegovac, Marija and Costello, Eithne and Greenhalf, William and Saad, Amina and Roberts, Rhiannon and others},
  journal={PLoS Medicine},
  volume={17},
  number={12},
  pages={e1003489},
  year={2020},
  publisher={Public Library of Science San Francisco, CA USA}
}

@article{yu2019maximum,
  title={Maximum smoothed likelihood component density estimation in mixture models with known mixing proportions},
  author={Yu, Tao and Li, Pengfei and Qin, Jing},
  journal={Electronic Journal of Statistics},
  volume={13},
  pages={4035--4078},
  year={2019}
}

@book{eggermont2001maximum,
  author = {Eggermont, P. P. B. and Lariccia, V. N.},
  title = {Maximum Penalized Likelihood Estimation},
  year = {2001},
  address = {New York},
  publisher = {Springer}
}

@book{pepe2003statistical,
  author = {Pepe, M. S.},
  title = {The Statistical Evaluation of Medical Tests for Classification and Prediction},
  year = {2003},
  address = {New York},
  publisher = {Oxford University Press}
}

@article{wand1994multivariate,
  title={Multivariate plug-in bandwidth selection},
  author={Wand, Matt P and Jones, M Chris and others},
  journal={Computational Statistics},
  volume={9},
  number={2},
  pages={97--116},
  year={1994},
  publisher={Heidelberg: Physica-Verlag,[1992-}
}

@article{ma2007combining,
  title={Combining multiple markers for classification using ROC},
  author={Ma, Shuangge and Huang, Jian},
  journal={Biometrics},
  volume={63},
  number={3},
  pages={751--757},
  year={2007},
  publisher={Oxford University Press}
}

\end{document}